\documentstyle[editedvolume,epsf]{crckapb}     

\def\/{$\,$}
\def\etal{{\it et al.}}
\def\ltsim{\mathrel{\hbox{\rlap{\hbox{\lower4pt\hbox{$\sim$}}}\hbox{$<$}}}}
\def\gtsim{\mathrel{\hbox{\rlap{\hbox{\lower4pt\hbox{$\sim$}}}\hbox{$>$}}}}
\def\hi{H$\>$I}
\def\msun{$M_{\odot}$}

\begin{opening}
\title{OVERVIEW:  LOW-z OBSERVATIONS}
\subtitle{Of Interacting and Merging Galaxies}
\author{FRANCOIS SCHWEIZER}
\institute{Carnegie Institution of Washington\\
           Department of Terrestrial Magnetism\\
           Washington,  DC 20015, USA}
\end{opening}

\runningtitle{OVERVIEW: LOW-z OBSERVATIONS}

\begin{document}

\begin{abstract}
Gravitational interactions and mergers affect the morphologies and dynamics of
galaxies from our Local Group to the limits of the observable universe.
Observations of interacting galaxies at low redshifts ($z\ltsim 0.2$) yield
detailed information about many of the processes at work.  I briefly review
these processes and the growing evidence that mergers play a major role in
the delayed formation of elliptical and early-type disk galaxies both in the
field and in clusters.  Low-$z$ observations clearly contradict the notion
of a single epoch of E formation at $z\gtsim 2$; instead, E and S0 galaxies
continue forming to the present.  The different rates of E and S0 formation
inferred from observations of distant and nearby clusters may partially
reflect the dependence of dynamical friction on mass: Major, E-forming
mergers may tend to occur earlier than minor, S0-forming mergers because the
dynamical friction is strongest for equal-mass galaxies.
\end{abstract}

\section{Introduction}

Three published figures illustrate the dramatic progress in our understanding
of galaxy structure and evolution made during the past six decades.  The
first of these is Hubble's (1936) empirical ``Sequence of Nebular Types''
(tuning-fork diagram), which begged the question: What determines
the position of a galaxy along this sequence?  And, more specifically, why
are galaxies at one end of the sequence disk-shaped and at the other end
ellipsoidal?  The second figure is Toomre's (1977) sketch of ``Eleven NGC
Prospects for Ongoing Mergers,'' which illustrated how this puzzling shape
dichotomy might result from disk galaxies merging to produce ellipticals
(Toomre \& Toomre 1972, hereafter TT for short).  The third figure is
Barnes's (1992, Fig.~9) display of the simulated final passages and merger
of two disk galaxies, each represented with its own live halo, disk, and
bulge.  This figure not only validates TT's earlier hypothesis, but also
shows that ellipticals formed via disk mergers bear signatures of the former
disks in their fine structure.

The early work by TT and others on gravitational interactions led to many
of the broader questions that preoccupy us at present:  What dynamical
processes drive galaxy evolution?  What is the relative importance of
tidal interactions, minor accretions, and major mergers in shaping the
various types of galaxies?  What fractions of stars formed quiescently versus
in violent episodes (Schechter \& Dressler 1987)?  How fast do galaxies
assemble?  And how has the mix of galaxy types evolved over time?

Although we should not expect this Symposium to yield definitive answers
to all these questions, the subject of galaxy interactions is progressing
rapidly at present, and the moment seems opportune to review recent advances
and chart new courses for addressing these questions.

\section{Tidal Interactions}

In essence, the tidal nature of galactic ``bridges'' and ``tails'' was
deciphered during the early 1970s (e.g., TT; for a review, see Barnes \&
Hernquist 1992). Because of our human preference
for the spectacular, there has been a strong observational bias toward
studies of near-equal-mass collisions and mergers.  Therefore, our knowledge
of unequal-mass collisions (say, $m/M\ltsim 0.3$) and of the cumulative
effects of weak, relatively distant interactions remains fragmentary.  For
example, of many galaxy deformations observed in the Local Group, only
perhaps the Magellanic Stream (e.g., Gardiner \& Noguchi 1996; Lin \etal\
1995) and the elongated dwarf galaxy Sgr~I (see below) are reasonably well
understood.  We still do not know the exact causes of the Milky Way's
and M33's warps, M31's misaligned bulge, or NGC 205's tidal deformation.

Yet, the eight years since the Heidelberg meeting (Wielen 1990) have brought
considerable progress in our general understanding of tidal interactions.
Foremost perhaps is a steadily growing appreciation of the many phenomena
associated with gas transport and induced star formation, as discussed 
briefly in \S 5 below and at length in the reviews by Kennicutt, Schweizer,
\& Barnes (1998).  And certainly a high point has been the discovery of the
Sagittarius dwarf, hitherto hidden behind the Milky Way's bulge and the
first clear case of accretion observed in our own galaxy (Ibata \etal\ 1995).
This apparently disintegrating companion strongly supports the hypothesis
that our halo formed gradually from accreting fragments (Searle \& Zinn 1978).

The ability of the {\it Very Large Array} to now routinely map the \hi\
kinematics of interacting galaxies is rejuvenating the study of these
systems.  Because tidal bridges and tails form from the gas-rich outskirts
of disk galaxies, they contain \hi\ along their full optically visible
extent and often significantly beyond (Hibbard \etal\ 1994;
Hibbard \& van Gorkom 1996).  The resulting velocity maps contain
a wealth of optically inaccessible information and yield kinematic
constraints that are invaluable for modeling these systems through
$N$-body simulations (e.g., Hibbard \& Mihos 1995).  Even the
tail lengths alone constrain the ratio of dark to luminous matter in
disk galaxies to $M_{\rm d}/M_{\rm l}\ltsim 10$ (Dubinski \etal\
1996; Mihos \etal\ 1998).  In the long run, detailed modeling of
interacting systems with mapped \hi\ kinematics should yield not only
this ratio for individual galaxies, but also the radial variation of it.

Evidence continues to grow that some dwarf galaxies form in tidal tails,
as originally proposed by Zwicky (1956).  Major clumps of stars and gas
have been found in many tails by now (Mirabel \etal\ 1991; Duc \& Mirabel
1994; Hunsberger \etal\ 1996), and two of these clumps have been shown to
probably be self-gravitating entities from their measured \hi\ velocity
dispersions (Hibbard \etal\ 1994).

Many issues remain to be addressed.  For example, despite assiduous work by
observers and numerical simulators alike, we still do not have any fully
successful models for the tidally generated spiral structures of M51 and
M81.  A new puzzle are the observed displacements between the stars and
\hi\ in some tidal tails, occasionally exceeding 2 kpc and perhaps
reflecting non-gravitational forces acting upon the gas (Hibbard \& van
Gorkom 1996; Schiminovich \etal\ 1995). Finally, although there are
many new observations of collisional ring galaxies (for a review,
see Appleton \& Struck-Marcell 1996), none have addressed yet the interesting
issue of the nature of ring-galaxy remnants: Into what kind of galaxies
do they evolve?

\section{Mergers and the Formation of Ellipticals}

The essence of this subject can be encapsulated in the following three
questions:  (1) What fraction of elliptical galaxies formed via major
disk--disk (DD) mergers?  (2) Did cluster ellipticals form via such DD
mergers, via multiple minor mergers, or in a single collapse?  And (3),
what is the age distribution of elliptical galaxies?


There is now strong evidence that at least some DD-merger remnants are
present-day protoellipticals, as envisaged by TT.  My own two favorites are
NGC 3921 and NGC 7252.  Both feature double tidal tails, but single main
bodies of $M_V\approx -23$ with $r^{1/4}$-type light distributions
indicative of violent relaxation (Schweizer 1996, 1982). Their power-law
cores, central luminosity densities, and $U\!BV\!I$ color gradients are
typical of ellipticals, and their central velocity dispersions fit the
Faber--Jackson relation for E's well (Lake \& Dressler 1986).
Their ``E\/+\/A'' spectra indicate recent ($\ltsim$1 Gyr ago) starbursts
of strength $b\approx 10$\%--30\%.  During these major starbursts, the
globular-cluster populations appear to have increased by $\gtsim$40\% in
NGC 3921 (Schweizer \etal\ 1996) and $\sim$80\% in NGC 7252 (Miller \etal\
1997).  Within 5--7 Gyr, both remnants will have specific globular-cluster
frequencies typical of field ellipticals.  Therefore, in all their observed
properties these two remnants appear to be 0.5--1 Gyr old protoellipticals.

A notable success has been the observational confirmation of the
theoretical prediction by Barnes (1988) that most of the tidally ejected
material must fall back. This phenomenon
creates a strong connection between DD mergers and field ellipticals.
In NGC 7252, and likely also in NGC 3921, the \hi\ in the lower parts of
the tails is observed returning toward the central remnant (Hibbard
\etal\ 1994; Hibbard \& Mihos 1995; Hibbard \& van Gorkom 1996).
It seems hardly coincidental, then, that many E and S0 galaxies feature
inclined gas disks (van Gorkom \& Schiminovich 1997), \hi\ absorption in
radio ellipticals always indicates infall (van Gorkom \etal\ 1989), and
some well-known E and S0 galaxies like NGC 5128, NGC 1052, and NGC 5266
possess {\it two}, often nearly orthogonally rotating, \hi\ disks
(Schiminovich \etal\ 1994; Plana \& Boulesteix 1996; Morganti \etal\ 1997).
There can be little doubt that at least these kinds of field galaxies
formed through major DD mergers.

Observations of fine structure in field E and S0 galaxies suggest that not
just a few, but {\it most} of these galaxies formed through that same
mechanism (Schweizer \& Seitzer 1992).  Roughly 70\% of the E's and
over 50\% of the S0's show fine structure (mainly ripples and plumes)
indicative of past disk mergers, and recent $N$-body simulations demonstrate
that this fine structure is a natural byproduct of tidal material falling
back after a major merger (Hernquist \& Spergel 1992; Hibbard \& Mihos 1995). 
Photometry of these structures suggests that there is more luminous matter
in them than can be accounted for by the proverbial ``gas-rich dwarf'' that
supposedly fell in (Prieur 1990).  Therefore, given the high detection rates
of fine structure and the limited duration of a significant flux of
returning material ($\ltsim$5 Gyr), it seems now likely that the vast
majority, and perhaps all, field E and S0 galaxies formed through DD mergers.  

For cluster ellipticals the situation is less clear because they are
poorer in \hi\ and fine structure.  Yet, there are at least three arguments
to support the view that these ellipticals formed through DD mergers
as well.  First, cluster ellipticals are structurally indistinguishable from
field ellipticals, for which the evidence for past DD mergers is strong.
Second, many cluster ellipticals possess oddly rotating stellar cores,
which seem to form naturally in DD mergers (Hernquist \& Barnes 1991;
Bender 1996) and point toward two, rather than multiple, merged components.
And third, the bimodal color distributions of globular clusters in cluster
ellipticals like M87 and M49 (Whitmore \etal\ 1995; Geisler \etal\ 1996)
imply a second major cluster-forming event, probably a merger (Ashman \&
Zepf 1992).  Although the remarkably small scatter in the color--luminosity
relations of cluster E and S0 galaxies is often taken as evidence against
major DD mergers, it may simply be a natural consequence of such mergers
having taken place in clusters relatively early (see \S 6).

\begin{figure}[t]    
\parbox[b]{6.2 cm}{\centering \leavevmode
\epsfxsize=0.49\columnwidth \epsfbox{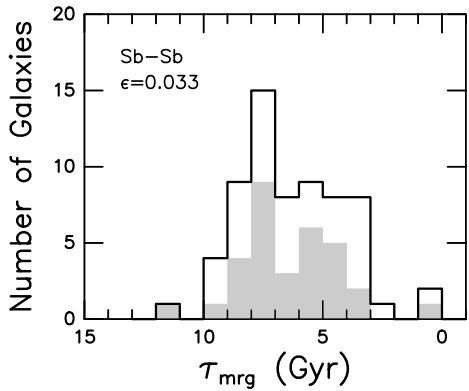}}\hfill
\parbox[b]{5.5 cm}{
\caption[]{Distribution of merger ages for 65 field E\/+\/S0 galaxies.
Ellipticals are shaded, age ``0'' marks present time.  Ages are computed for
presumed Sb--Sb mergers with star-formation efficiency $\epsilon${}\/=\/0.033.
Note prolonged period of E\/+\/S0 formation suggested by these ages.  After
Schweizer \& Seitzer (1992). \vspace{3mm}}
}
\end{figure}

If ellipticals did form through delayed DD mergers (Toomre 1977),
their formation ages should show a wide spread.  The evidence for a
significant age spread is increasing, despite the fact that ages of
individual E's remain uncertain and often controversial (O'Connell 1980,
1994; Schweizer \& Seitzer 1992; Gonz\'alez 1993; Faber \etal\ 1995; Davies
1996).  One problem lies in the different possible definitions of ``age.''
In an elliptical galaxy, one must distinguish at least three ages:
(1) the true mean age $\langle\tau_{\ast}\rangle$ of the stars,
(2) the luminosity-weighted mean age $\langle\tau_{\ast}\rangle_{\rm lum}$
of stars, which is what observers using {\it single}-burst population models
for interpretation typically measure, and
(3) the merger age $\tau_{\rm mrg}$, reckoned since the merger began or the
starburst peaked. In general,
$\langle\tau_{\ast}\rangle > \langle\tau_{\ast}\rangle_{\rm lum} >
\tau_{\rm mrg}$.  Figure 1 shows merger ages for 65 E and S0 galaxies
based on $U\!BV\!$ colors and a simple \mbox{{\it two}-burst} model of star
formation. These formation ages spread over most of the age of the universe.
The above model and Kauffmann's (1996) similar calculations also
clarify why age-dating ellipticals from spectra is difficult: Because
the bulk of stars formed before the final mergers, ellipticals {\it look}
more uniform and old than they really are.

In summary, the combined evidence from low-$z$ observations strongly indicates
that {\it there was no single ``epoch of E formation'' at $z\gtsim2$}, as
some have inferred from high-$z$ observations. Instead, this evidence
favors the view that E and S0 galaxies continue forming to the present
and will do so into the future.

\section{Mergers in Disk Galaxies}

Although equal-mass mergers are the most spectacular, there must also be
unequal-mass (``minor'') mergers that affect disk galaxies without
completely destroying their disks.  Three central questions are: (1) Can
bulges form through minor mergers?  (2) What fraction of bulges formed in
this manner?  And (3) how fragile are galaxy disks?

Theoretical work on the fragility of disks is undergoing rapid revisions.
As recently as 1992, T\'oth \& Ostriker argued that disks are very fragile
and would be disrupted by any infalling companion more than a few percent
the mass of the main disk.  Yet, $N$-body simulations suggest that disk
galaxies can survive minor mergers of up to $m/M\approx 0.3$, albeit with
increases in bulge mass and a thickened disk (Walker \etal\ 1996).

Recent observations suggest that the effects of minor mergers on disk
galaxies can be surprisingly complex.  For example, S0 galaxies with polar
rings were thought to have accreted their ring gas during a merger.
Yet, new observations show that the \hi\ content of polar rings is often
large and typical of late-type disk galaxies, and that many of the S0's
feature poststarburst spectra (e.g., Richter \etal\ 1994; Reshetnikov \&
Combes 1994).  Thus it looks as if it is the central S0 galaxies (rather
than the polar rings) that may have formed from a gas-rich companion falling
into a spiral galaxy nearly over its poles.  If so, these central S0 bodies
represent {\it failed bulges} (Schweizer 1995; Arnaboldi \etal\ 1997).

Various observations suggest that quite in general early-type
disk galaxies may be the remnants of minor mergers.  From the
statistics of counterrotating, skewedly rotating, and corotating
ionized gas disks, one can conclude that 40\%--70\% of all S0 galaxies
experienced minor mergers (Bertola \etal\ 1992).  The phenomenon of
counterrotating gas disks is observed---with decreasing frequency---into
Hubble types S0/a, Sa, and Sab.

Another powerful kinematic signature of past mergers are subpopulations of
stars counterrotating in disk galaxies of types S0\/--\/Sb.   The best
known example of this phenomenon occurs in NGC 4550, an E/S0
galaxy with half of its disk stars rotating one way and the other half the
other way (Rubin \etal\ 1992).  In the Sa galaxy NGC 4138, the split between
normal- and counterrotating disk stars is 75/25\%  (Jore \etal\ 1996),
while in the Sb galaxy NGC 7217 it is 70/30\% (Kuijken 1993).  Finally,
the whole bulge seems to counterrotate to the disk in the well-known
Sb galaxy NGC 7331 (Prada \etal\ 1996).  Some first $N$-body simulations
suggest that minor and not-so-minor dD mergers can indeed produce the
counterrotations observed in these systems (Thakar \etal\ 1997;
Pfenniger 1998).

A connection between mergers in disk galaxies and bulge formation is also
suggested by fine structure indicative of past mergers observed in many S0
and Sa galaxies (Schweizer \& Seitzer 1988). Even NGC 4594, the ``Sombrero,''
sports a faint fan of luminous material and an opposite tail signaling a not
too ancient merger (Malin \& Hadley 1997).  Our present understanding of the
effects of mergers on disk galaxies is then as follows.

Minor mergers do occur in disk galaxies and seem to move them toward
earlier Hubble types.  It appears that disks, especially those with
a significant fraction of gas, are not nearly as fragile as thought
just a few years ago.  However, we must remember that bulgeless
(e.g., M33) and lopsided disk galaxies do constrain the rate of minor
mergers. For mergers of $m/M\approx 0.1$, this rate is estimated to be
$\ltsim$0.07\/--\/0.25 events/Gyr (Zaritsky \& Rix 1997).  At present,
we can merely state that the fraction of bulges built through mergers
is clearly $>$0, but its value remains unknown.  A challenge for the
future is to determine how unique or varied the possible paths to, say,
a present-day Sb galaxy are.  Does the disk form first or the bulge, and
can each grow episodically and perhaps even by turns?

\section{Interaction-Induced Processes}

The {\it IRAS\/} sky survey opened our eyes to the fact that gas plays
a disproportionately large role in interacting and merging galaxies.
Due to lack of space, I can here only briefly sketch recent progress in
our understanding of the four major interaction-induced processes.

Interaction-induced {\it starbursts\/} are fierce episodes of ``galaxy
building,'' during which 5\%--20\% of the luminous matter gets
converted from gas into stars over periods of $\sim$10$^8$ years.
Fueled by molecular gas in quantities of up to
$\sim$2$\,\times${}10$^{10}$\/\msun, these starbursts appear to be
self-limiting: the
star-formation rate does not exceed $\sim$0.7\/\msun\/kpc$^{-2}$\/yr$^{-1}$
for stars of 5--100\/\msun.  Merger-induced torques tend to drive gas
toward the center, where high concentrations of it often dominate the
dynamics.  These induced concentrations may explain the high central
phase-space densities observed in E's.

{\it Globular-cluster formation\/} appears to be a natural by-product of
induced starbursts (e.g., NGC 4038/39, 7727, 3921, 7252, 5128, 1275).
There is growing evidence that the globulars may form preferentially in
high-pressure regions from Giant Molecular Clouds shocked by the
surrounding starburst-heated gas.  Even at $z\approx 0$, gas-rich
DD mergers can apparently about double the number of globular clusters.

{\it Galactic winds\/} are radial $10^2$--$10^3$\/km\/s$^{-1}$ outflows of
gas heated mostly by starburst supernovae.  These winds appear
``mass loaded'' by factors of 3--6.  In M82, the Fe-rich wind is estimated
to eject $\sim$10$^8$\/\msun\ of gas into the halo over 20--30\/Myr.
Such winds presumably played a major role in the chemical evolution of
early-type galaxies, but details remain unclear.

{\it Nuclear activity\/} often accompanies merger-induced gas inflows.
Observationally, assessing the relative contributions from active
galactic nuclei and central starbursts remains a challenging task.
About 80\% of all host galaxies of low-$z$ QSOs appear to be interacting
and 30\% appear to be merging.  There is evidence that the QSO activity
tends to peak shortly before the nuclei of two galaxies merge, and in at
least the case of OX\/169 the variable H$\beta$ emission-line components
suggest the presence of {\it two\/} separate engines (Stockton \&
Farnham 1991).

\section{Interactions and Mergers in Clusters}

Clusters are complex environments for galaxy evolution.  Two properties
of cluster galaxies have led to major questions.  First, is the
morphology--density relation (Oemler 1974; Dressler 1980) a result of
birth or evolution?  And second, is the Butcher-Oemler (1978, 1984)
effect mainly due to ram-pressure stripping or interactions and mergers?
Many astronomers have doubted whether galaxies in clusters can merge
because of their high mean relative velocities.  If not, why are
there so many E\/+\/S0 galaxies in clusters?

Yet, there has long been observational evidence for interactions and
mergers in clusters, and recently this evidence has grown stronger.

For example, the Hercules Cluster abounds in pairs of interacting disks.
The Coma Cluster harbors the well-known DD merger NGC 4676 (TT; Barnes
1998) in its outskirts. And in the Virgo Cluster, systems like NGC 4438/35
show that even $\sim$10$^3$\/km\/s$^{-1}$ collisions can strongly affect
member galaxies (Combes \etal\ 1988; Kenney \etal\ 1995). Clearly, strong
tidal interactions and mergers occur in clusters to the present time.

Mergers must have played a major role in shaping the E and S0 galaxies
of the Virgo Cluster, especially the most massive ones (as judged by their
having Messier numbers).  Many of these galaxies feature oddly rotating
subsystems (M86, NGC 4365, NGC 4550), ripples (M85, M89), and bimodal
globular-cluster populations (M49, M87), all signatures of past
disk mergers.  This confirms predictions of $N$-body simulations
which have long suggested that ``the upper end of the
mass spectrum is most strongly affected by the merging process occurring
predominantly during the expansion and subsequent collapse of the cluster''
(Roos \& Aarseth 1982).

There is also growing evidence for interactions and mergers in clusters
at $z\approx 0.2$--0.5.  Even before the advent of {\it HST}, groundbased
observations suggested that many blue galaxies in Butcher-Oemler clusters
are distorted disks or feature excess companions (Lavery \& Henry 1988--1994).
Observations with {\it HST\/} show that a fair fraction of these galaxies
are interacting or merging while a majority appear to be disturbed gas-rich
disks (Dressler \etal\ 1994; Couch \etal\ 1994; Barger \etal\ 1996).
Perhaps the most interesting recent result is that at $z\approx 0.4$
the E population appears to be fully in place, while there are
significantly fewer S0 galaxies and more late-type spirals and Irregulars
than at $z\approx 0$ (Oemler \etal\ 1997).

I believe that a viable scenario for the formation of E and S0 galaxies
in clusters is as follows.  Major DD mergers occurred relatively early
because the dynamical friction is strongest for equal-mass galaxies.
These mergers formed the bulk of the ellipticals.  Minor mergers took
longer on average because of their lesser dynamical friction (deceleration
being approximately proportional to mass). They---and perhaps also multiple
interactions---slowly transform spirals into S0 galaxies. Presumably, it is
the evolving substructure of clusters that allows mergers to occur to the
present time.

Finally, field galaxies experienced the same processes, but on a longer
time scale because of their lower spatial density.  This is why we
still see ellipticals forming (NGC 3921, NGC 7252, ULIR galaxies) and
why there are fewer S0 galaxies and more spirals in the field than in
clusters.  If this cluster- and field-evolution scenario is correct in its
essence, Hubble's morphological sequence may rank galaxies mainly by
the number and mass ratio of mergers in their past history.

I gratefully acknowledge support through NSF Grant AST-95\/29263.


\end{document}